\documentclass[a4paper,11pt,preprintnumbers]{article}
\pdfoutput=1

\usepackage{jcappub}
\usepackage[utf8]{inputenc}
\usepackage{booktabs}
\usepackage{graphicx,multirow,amsmath,amsfonts,amssymb}
\usepackage[dvipsnames]{xcolor}
\usepackage{hyperref}
\usepackage{float}
\usepackage{mathtools}
\usepackage{enumitem}
\usepackage{tcolorbox}
\tcbuselibrary{skins,breakable}

\newcommand{\sw}{\sin^2\!\theta_w}
\newcommand{\nuc}{{\rm nuc}}
\newcommand{\ap}{\alpha'}
\newcommand{\astr}{\alpha_{\rm str}}
\newcommand{\kIKT}{\kappa}
\newcommand{\mstr}{\mu_{\rm str}}
\newcommand{\Tnuc}{T_{\rm nuc}}

\newcommand{\SB}{S_B}
\newcommand{\flow}{f_{\rm low}}
\newcommand{\fstar}{f_{*}}
\newcommand{\Gmu}{G_N\mu_{\rm str}}
\newcommand{\MPl}{M_{\rm Pl}}

\newcommand{\be}{\begin{equation}}
\newcommand{\ee}{\end{equation}}
\newcommand{\rev}[1]{#1}
\newcommand{\IPPP}{Institute for Particle Physics Phenomenology, 
  Durham University, South Road, DH1 3LE, Durham, United Kingdom}

\definecolor{insightcol}{RGB}{0,80,40}
\tcbset{insightbox/.style={colframe=insightcol,colback=insightcol!6,
  arc=3pt,boxrule=1pt,breakable}}

\title{\huge Thermal Metastable Strings in One-Scale Models and Gravitational Waves}

\author[a]{Arturo de Giorgi,}
\author[a]{James Ingoldby,}
\author[a]{Valentin V.\ Khoze,}
\author[a]{and Jessica Turner\,}

\affiliation[a]{\IPPP}

\emailAdd{arturo.de-giorgi@durham.ac.uk}
\emailAdd{james.a.ingoldby@durham.ac.uk}
\emailAdd{valya.khoze@durham.ac.uk}
\emailAdd{jessica.turner@durham.ac.uk}

\abstract{
Metastable cosmic strings provide a cosmological interpretation of the nanohertz
stochastic gravitational wave background reported by Pulsar Timing Array (PTA)
experiments. We revisit this scenario in a minimal dark-sector gauge theory, in
which a complex Higgs doublet breaks $\mathrm{SU}(2)\times\mathrm{U}(1)\to\mathrm{U}(1)_{\rm IR}$
at a single symmetry-breaking scale. This one-scale setup predicts metastable
$Z$-strings whose endpoints are monopole-like defects, and whose zero-temperature
decay rate is controlled by the gauge couplings and mass ratios. We show that,
once the string-forming transition occurs in a thermal plasma, the dominant decay
channel is not the zero-temperature monopole nucleation but thermally induced
nucleation on the string worldsheet. We determine the nucleation temperature
$T_{\rm nuc}$, which we \rev{adopt as the reference string-formation scale}, from the
one-loop finite-temperature effective potential with daisy resummation, and use it
to evaluate the worldsheet bounce action throughout the model parameter space.
\rev{Requiring a PTA-compatible gravitational wave}
signal selects a narrow region in the model parameter space, in the
$(\sin^2\theta_w,\sqrt{\beta})$ plane, where $\theta_w$ is the dark-sector weak
mixing angle and $\beta\equiv M_\Phi^2/M_{Z}^2$ is the squared Higgs-to-$Z$ mass
ratio. Thermal effects modify the zero-temperature picture significantly, shifting
the PTA-compatible region towards lower values of the dark fine-structure constant
$\alpha'$ and larger values of the monopole-to-string-tension ratio $\kappa$.
}


\begin{document}
\maketitle

%%%==========================================================================
\section{Introduction}
\label{sec:intro}
%%%==========================================================================
The nanohertz stochastic gravitational wave background reported by Pulsar
Timing Array (PTA) collaborations
\cite{NANOGrav:2023gor,NANOGrav:2023icp,NANOGrav:2023hfp,
NANOGrav:2023ctt,NANOGrav:2023hvm,EPTA:2023fyk,EPTA:2023sfo,
EPTA:2023akd,EPTA:2023gyr,EPTA:2023xxk,
EuropeanPulsarTimingArray:2023egv,Zic:2023gta,Reardon:2023zen,
Reardon:2023gzh,Xu:2023wog}
has sparked interest in cosmic strings as a possible cosmological source of the signal. Stable Nambu--Goto string networks are, however, strongly constrained by the same data. This has motivated particular attention to
metastable strings, whose finite lifetime modifies the gravitational wave spectrum produced by the network. Metastable strings can break by nucleating monopole--antimonopole pairs on the string worldsheet~\cite{Preskill:1992ck}. The resulting decay of the network suppresses the spectrum at low frequencies~\cite{Buchmuller:2020lbh,Buchmuller:2021mbb}, providing a natural mechanism for producing a PTA-band signal while avoiding the excessive low-frequency power associated with stable string networks.

Most realisations of metastable strings are based on two-scale symmetry-breaking patterns. The minimal example is
$\mathrm{SU}(2)\to\mathrm{U}(1)\to\emptyset$, in which monopoles of mass $M_m$ are produced at the first stage of spontaneous symmetry breaking, and strings of tension $\mu_{\rm str}$ form at the second. This structure is well motivated by grand-unified model building, since many GUT-breaking chains pass through intermediate gauge symmetries that realise this pattern~\cite{Buchmuller:2019gfy,Fu:2023mdu,Antusch:2023zjk,Antusch:2024nqg}.
In such models, the lifetime of the string network is controlled by the microscopic ratio
\begin{equation}
  \kappa \equiv \frac{M_m^2}{\mu_{\rm str}}\,.
\end{equation}
At zero temperature, the breaking rate per unit length takes the form
\cite{Preskill:1992ck,Shifman:2002yi,Monin:2008mp,Monin:2009ch,Leblond:2009fq,Ingoldby:2025wcl}
\begin{equation}
  \Gamma \simeq \frac{\mu_{\rm str}}{2\pi}\,e^{-\pi\kappa}\,,
\end{equation}
so that the hierarchy between the monopole mass and the string tension exponentially controls the lifetime of the network. %In this cold-formation picture\rev{, realised when the network forms outside a thermal bath --- for instance in a waterfall transition at the end of hybrid inflation~\cite{Linde:1993cn,Copeland:1994vg,Buchmuller:2012wn,Buchmuller:2013lra} ---} fits to PTA-motivated spectra typically select $\sqrt{\kappa}\simeq 7$--$9$, subject to assumptions about loop emission, segment emission, and the time at which the network collapses~\cite{Buchmuller:2020lbh,Buchmuller:2021mbb,Antusch:2023zjk,Madge:2023dxc}. 
\rev{This cold-formation picture is valid whenever the network forms outside a thermal bath, for instance in a waterfall transition at the end of hybrid inflation~\cite{Linde:1993cn,Copeland:1994vg,Buchmuller:2012wn,Buchmuller:2013lra}. The fit to PTA-motivated spectra then typically selects $\sqrt{\kappa}\simeq 7$--$9$, subject to assumptions about loop emission, segment emission, and the time at which the network collapses~\cite{Buchmuller:2020lbh,Buchmuller:2021mbb,Antusch:2023zjk,Madge:2023dxc}.}
\rev{The classical stability of the strings themselves is a further requirement: in the commonly studied two-step setup~\cite{Chitose:2023dam}, instabilities have recently been shown to affect parts of the PTA-relevant parameter space~\cite{Blasi:2026iyq}.}

There is, however, another possible mechanism by which unstable string networks can be terminated. In the quasi-stable string scenario of Refs.~\cite{Lazarides:2022jgr,Lazarides:2023ksx,Maji:2026nkz}, monopoles are produced at an earlier stage in the cosmological history, and the subsequent string-forming transition connects monopoles and antimonopoles by finite string segments. If these segments are much longer than the horizon at formation, the network evolves like an ordinary long-string network for the purpose of gravitational wave emission. Its eventual disappearance is not controlled by the nucleation of new monopole pairs on the worldsheet, but by the later horizon re-entry of the pre-existing endpoints, which are then pulled together by the string tension.

\rev{At finite temperature, these two possibilities merge: a metastable string network can realise the quasi-stable scenario dynamically~\cite{Tranchedone:2026lav}.} In two-scale models, the high temperature of the string-forming transition exponentially enhances the rate at which monopoles nucleate on strings. An initially long-string network can therefore be chopped into super-horizon segments shortly after formation, while the subsequent monopole nucleation rate becomes highly suppressed as the Universe cools. The resulting long-lived segments then realise a quasi-stable string network. In addition, strings can attach to thermally produced monopoles during network percolation, providing another route to the same late-time cosmology. Ref.~\cite{Asl:2026zpj} developed a unified gravitational wave template for these different possibilities by distinguishing the microscopic string-breaking time from the later network-collapse time. \rev{Determining whether a given microscopic string model can account for the PTA signal therefore requires tracking three ingredients: the classical stability of the string, the finite-temperature corrections to its decay rate, and the cosmological history of the string network.}

\rev{In Ref.~\cite{Ingoldby:2025wcl}, we addressed the first two of these ingredients at zero temperature, for a complementary possibility:}
metastable strings in a one-scale dark sector. The model is an
electroweak-like $\mathrm{SU}(2)\times\mathrm{U}(1)$ gauge theory broken to a
residual $\mathrm{U}(1)_{\rm IR}$ by a single Higgs doublet,
$\mathrm{SU}(2)\times\mathrm{U}(1)\to\mathrm{U}(1)_{\rm IR}$. This symmetry-breaking
pattern admits embedded $Z$-string solutions, analogous to the electroweak
strings first constructed in Ref.~\cite{Vachaspati:1992fi}. Although such
strings are not topologically protected, they can be dynamically stable in
the near-semi-local regime
\cite{Vachaspati:1991dz,Hindmarsh:1991jq,James:1992wb,Achucarro:1999it}.
In the dark-sector realisation of Ref.~\cite{Ingoldby:2025wcl}, the
corresponding classically stable $Z$-strings decay semiclassically by nucleating monopole-like endpoints. We computed the string tension by solving the Nielsen--Olesen
equations~\cite{Abrikosov:1956sx,Nielsen:1973cs}, determined the monopole mass in the thin-defect approximation, and mapped the Lagrangian parameters
onto the zero-temperature decay parameter $\kappa$. This showed that a
classically stable region of the one-scale model can reproduce the
zero-temperature PTA window without invoking extended Higgs sectors or
multi-stage symmetry breaking.

In this paper, we revisit the same one-scale construction and include the finite-temperature phase transition at which the string network forms\rev{, together with the resulting network history. We will focus on the case of a first-order phase transition as this allows us to connect the Lagrangian parameters with the string formation temperature and cosmological history.}. 
This changes the interpretation of the microscopic parameters. The three-dimensional thermal action $S_3(T)$ determines when bubbles of the
broken phase nucleate and hence when the string network forms. In contrast,
the worldsheet bounce action $S_B(T)$ determines the rate at which the newly formed
strings break into segments by monopole--antimonopole nucleation. These two actions describe different processes, but they are linked because the string-breaking action must be evaluated at the temperature selected by the phase transition\rev{: the formation history fixes the PTA condition. For cold formation, as in the waterfall scenario above, the worldsheet action receives no thermal correction and retains its vacuum value $S_B(0)=\pi\kappa$; the cold-formation analysis of Ref.~\cite{Ingoldby:2025wcl} then applies. For the hot dark-sector transition studied here, the PTA requirement is instead imposed on $S_B(T_{\rm nuc})$, where $T_{\rm nuc}$ is the nucleation temperature of the transition, which we adopt as the reference string-formation scale: the low-frequency spectral cutoff $\fstar$ (defined in Section~\ref{sec:mechanism}) falls near $10^{-8}\,{\rm Hz}$ when $S_B(T_{\rm nuc})\simeq 109$, rather than at the zero-temperature condition $S_B(0)=\pi\kappa\simeq 200$ associated with $\sqrt{\kappa}\simeq 8$. The thermal and cold-waterfall cases are thus the two limiting formation histories of the same one-scale model, and comparing them quantifies how strongly the PTA-compatible parameter space depends on the early-Universe environment at the time of string formation.}

We carry out this analysis directly in the $\mathrm{SU}(2)\times\mathrm{U}(1)$ model. We compute the one-loop finite-temperature effective potential, including the Arnold--Espinosa daisy resummation of the bosonic Matsubara zero modes~\cite{Arnold:1992rz}, and use it to determine the nucleation temperature $T_{\rm nuc}$ of the string-forming transition. We then evaluate the finite-temperature worldsheet bounce action at this temperature and scan the microscopic parameter space spanned by the dark fine-structure constant $\alpha'$, the dark weak mixing angle $\theta_w$, and the squared Higgs-to-$Z$ mass ratio $\beta \equiv {M_\Phi^2}/{M_{Z}^2}$.
\rev{We find that the requirement of a PTA-compatible low-frequency cutoff selects a region in the $(\sin^2\theta_w,\sqrt{\beta})$ plane for}
$\alpha'\simeq 0.12$--$0.25$. Typical viable points have
\rev{$T_{\rm nuc}/\eta\simeq 0.4$--$0.6$} and $\kappa$ of order several hundred.
\rev{Finite-temperature effects change the one-scale analysis at a qualitative level, relocating the PTA-compatible region to parametrically different couplings: the viable window has smaller $\alpha'$ and substantially larger $\kappa$ than the cold-formation window identified in Ref.~\cite{Ingoldby:2025wcl}.} Physically, this corresponds to a regime in which the non-abelian factor is much more weakly coupled than the abelian one ($g \ll g'$).

% {\color{red} Isn't this paragraph a repetition? I would cut it entirely.}
% \rev{It is important to emphasise that this thermal formation scenario represents just one possible cosmological history. If, instead, the network forms outside a thermal bath (for instance via a cold waterfall transition driven by a coupling to the inflaton \cite{Linde:1993cn,Copeland:1994vg,Buchmuller:2012wn,Buchmuller:2013lra}) the worldsheet action receives no thermal correction and retains its vacuum value $S_B(0)=\pi\kappa$. In that limit, the cold-formation analysis of Ref.~\cite{Ingoldby:2025wcl} applies. The thermal scenario studied here and the cold waterfall scenario thus constitute the two limiting formation histories of the same one-scale model. By comparing the two, we can approximate how strongly the PTA-compatible parameter space depends on the early-Universe environment at the time of string formation.}

The paper is organised as follows.
Section~\ref{sec:model} reviews the one-scale
$\mathrm{SU}(2)\times\mathrm{U}(1)$ model and the zero-temperature quantities entering the string tension and monopole mass.
Section~\ref{sec:mechanism} reviews the finite-temperature
string-breaking action and the mechanism by which the broken network sources gravitational waves. The resulting signal depends crucially on the string-formation temperature, for which we adopt the phase transition nucleation temperature as a calculable reference scale.
Section~\ref{sec:thermal} describes the finite-temperature effective potential and the computation of \(T_{\rm nuc}\).
Section~\ref{sec:results} presents the parameter-space scan and the resulting PTA-compatible region.
Section~\ref{sec:conclusions} contains our conclusions.

%%%==========================================================================
\section{The Dark Electroweak Model and Parameters}
\label{sec:model}
%%%==========================================================================

We work in the electroweak-like one-scale model and use the notation of Ref.~\cite{Ingoldby:2025wcl}. The gauge group is $G=\mathrm{SU}(2)\times \mathrm{U}(1)$, with gauge couplings $g$ and $g'$, respectively. The scalar sector contains a single complex Higgs doublet $\Phi$, transforming as a fundamental under $\mathrm{SU}(2)$ and carrying
hypercharge $Y=1/2$. The Higgs field acquires a vacuum expectation value $\eta/\sqrt{2}$ and breaks the gauge symmetry in a single step:
\begin{equation}
  \mathrm{SU}(2)\times\mathrm{U}(1)\to \mathrm{U}(1)_{\rm IR}\,.
\end{equation}
This is a one-scale construction: the gauge-boson masses, the Higgs mass, the
string tension and the monopole endpoint energy are all controlled by the
same symmetry-breaking scale $\eta$.
The Lagrangian is
\begin{equation}
  \mathcal{L}
  =
  -\frac14 W^a_{\mu\nu}W^{a\mu\nu}
  -\frac14 B_{\mu\nu}B^{\mu\nu}
  + |D_\mu\Phi|^2
  -\lambda\left(\Phi^\dagger\Phi-\frac{\eta^2}{2}\right)^2 ,
  \label{eq:model_lagrangian}
\end{equation}
where $W^a_\mu$ with $a=1,2,3$ are the $\mathrm{SU}(2)$ gauge fields and
$B_\mu$ is the $\mathrm{U}(1)$ gauge field. The corresponding field strengths
are
\begin{equation}
  W^a_{\mu\nu}
  =
  \partial_\mu W^a_\nu-\partial_\nu W^a_\mu
  +g\,\epsilon^{abc}W^b_\mu W^c_\nu ,
  \qquad
  B_{\mu\nu}
  =
  \partial_\mu B_\nu-\partial_\nu B_\mu \,.
  \label{eq:field_strengths}
\end{equation}
The covariant derivative acting on $\Phi$ is
\begin{equation}
  D_\mu \Phi
  =
  \left(
    \partial_\mu
    -\frac{i g}{2}\tau^a W^a_\mu
    -\frac{i g'}{2}B_\mu
  \right)\Phi \,,
  \label{eq:covariant_derivative}
\end{equation}
where $\tau^a$ are the Pauli matrices. The parameter $\lambda$ is the Higgs
self-coupling, while $\eta$ sets the symmetry-breaking scale. In the broken
phase we take
\begin{equation}
  \langle\Phi\rangle
  =
  \frac{\eta}{\sqrt{2}}
  \begin{pmatrix}
    0\\
    1
  \end{pmatrix}\,,
  \qquad
  \langle\Phi^\dagger\Phi\rangle=\frac{\eta^2}{2}\,.
  \label{eq:higgs_vev}
\end{equation}
After symmetry breaking, the neutral gauge fields are
\begin{equation}
  Z_\mu=-\sin\theta_w\,B_\mu+\cos\theta_w\,W^3_\mu\,,
  \qquad
  A_\mu=\cos\theta_w\,B_\mu+\sin\theta_w\,W^3_\mu\,,
\end{equation}
where the dark Weinberg angle is defined by
\begin{equation}
  \sin^2\theta_w=\frac{g'^2}{g^2+g'^2}\,.
  \label{eq:sw_def}
\end{equation}
The massive gauge boson and Higgs masses are
\begin{equation}
  M_W=\frac{g\eta}{2}\,,
  \qquad
  M_Z=\frac{\eta}{2}\sqrt{g^2+g'^2}\,,
  \qquad
  M_\Phi=\sqrt{2\lambda}\,\eta \,.
  \label{eq:masses}
\end{equation}
We parametrise the scalar-to-vector mass ratio by
\begin{equation}
  \beta\equiv\frac{M_\Phi^2}{M_Z^2}\,.
  \label{eq:beta_def}
\end{equation}
The model admits embedded $Z$-string solutions. Their tension can be written
as
\begin{equation}
  \mu_{\rm str}=\eta^2\alpha_{\rm str}(\beta)\,,
  \label{eq:mu_str_def}
\end{equation}
where $\alpha_{\rm str}(\beta)$ is obtained by solving the Nielsen--Olesen profile equations, as in Ref.~\cite{Ingoldby:2025wcl}. The corresponding delocalised, large-radius flux configuration has tension
\begin{equation}
  \mu_\infty=\eta^2\alpha_\infty\,,
  \qquad
  \alpha_\infty=\pi \,,
  \label{eq:alpha_infty_def}
\end{equation}
and sets the energy scale of the monopole-like endpoints on which the
$Z$-string terminates, see Eq.~\eqref{eq:Mm_def} below.
For $\beta<1$, one has $\alpha_{\rm str}<\alpha_\infty$, so the localised string is energetically preferred over the delocalised configuration. In the near-semi-local regime, where $g\ll g'$ and $\beta<1$, the $Z$-string can be a classically stable local minimum of the energy functional. Since the vacuum
manifold is simply connected for non-zero $g$, however, the string is not
topologically protected. It can therefore decay semiclassically by nucleating monopole-like endpoints on the string.
In the thin-defect approximation, the monopole endpoint mass is
\begin{equation}
  M_m
  =
  \sqrt{8\pi\alpha_\infty}\,\frac{\eta}{g}
  =
  \sqrt{8\pi^2}\,\frac{\eta}{g}\,,
  \label{eq:Mm_def}
\end{equation}
where in the second equality we used $\alpha_\infty=\pi$. The corresponding zero-temperature decay parameter is
\begin{equation}
  \kappa
  \equiv
  \frac{M_m^2}{\mu_{\rm str}}
  =
  \frac{8\pi\alpha_\infty}{g^2\alpha_{\rm str}(\beta)}
  =
  \frac{8\pi^2}{g^2\alpha_{\rm str}(\beta)} \,.
  \label{eq:kappa_def}
\end{equation}
The zero-temperature string-breaking action is then
\begin{equation}
  S_B(0)=\pi\kappa \,.
  \label{eq:SB0_def}
\end{equation}
For the finite-temperature scan, it is convenient to trade $g$ and $g'$ for
the dark fine-structure constant $\alpha'\equiv g'^2/(4\pi)$ and
$\sin^2\theta_w$. From Eq.~\eqref{eq:sw_def}, one obtains
\begin{equation}
  g^2
  =
  4\pi\alpha'\,\frac{1-\sin^2\theta_w}{\sin^2\theta_w}\,,
  \label{eq:g_from_ap_sw}
\end{equation}
and therefore
\begin{equation}
  \kappa
  =
  \frac{2\pi\sin^2\theta_w}
       {\alpha'(1-\sin^2\theta_w)\alpha_{\rm str}(\beta)} \,.
  \label{eq:kappa_ap_sw}
\end{equation}
This relation is used throughout the scan to map the microscopic parameters $(\alpha',\sin^2\theta_w,\beta)$ onto the string-breaking parameter $\kappa$.

%%%==========================================================================
\section{Finite-temperature breaking mechanism}
\label{sec:mechanism}

Before turning to the detailed thermal calculation, we summarise the physical mechanism by which a network of metastable $Z$-strings produces a
PTA-band gravitational wave signal when the string-forming transition takes place in a thermal plasma and temperature effects are non-negligible. 
The picture follows the finite-temperature
analysis of Ref.~\cite{Tranchedone:2026lav}, specialised here to the one-scale $\mathrm{SU}(2)\times\mathrm{U}(1)$ model of Section~\ref{sec:model}. The gravitational wave signal is still sourced by loop emission during the quasi-scaling regime, as in the zero-temperature case. What differs is the cosmological history of the network itself, and in particular, the mechanism that ends its scaling evolution.

%^^^^^^^^^^^^^^^^^^^^^^^^^^^^^^^^^^^^^^
\subsection{Worldsheet bounce: zero and finite temperature}

The decay of the network is regulated by the breaking rate per unit string length $\Gamma$, given by the formula%~\cite{Preskill:1992ck,Monin:2008mp}
\begin{equation}
  \Gamma \;\simeq\; \frac{\mu_{\rm str}}{2\pi}\,e^{-S_B}\,,
  \label{eq:Gamma}
\end{equation}
where $S_B$ is the worldsheet bounce action. The temperature dependence of the prefactor is left implicit.

At zero temperature, a string breaks by quantum tunnelling: a circular Euclidean instanton of true vacuum, of radius $r_M = M_m/\mu_{\rm str}$,
is nucleated on the worldsheet. The corresponding bounce action and
breaking rate are~\cite{Preskill:1992ck}
\begin{equation}
  S_B(0) \;=\; \pi\,\frac{M_m^2}{\mu_{\rm str}} \;=\; \pi\kappa\,,
  \qquad
  \Gamma_0 \;\simeq\; \frac{\mu_{\rm str}}{2\pi}\,e^{-\pi\kappa}\,.
  \label{eq:Gamma0_mech}
\end{equation}
At finite temperature, the string worldsheet is compactified along the Euclidean time direction on a circle of period $1/T$. The vacuum bubble fits inside the compactification only for $T<T_0$, where
\begin{equation}
  T_0 \;\equiv\; \frac{\mu_{\rm str}}{2M_m}
      \;=\; \frac{1}{2}\sqrt{\frac{\mu_{\rm str}}{\kappa}}
      \;=\; \frac{\eta}{2}\sqrt{\frac{\alpha_{\rm str}}{\kappa}}.
  \label{eq:T0_mech}
\end{equation}
For $T>T_0$ the bubble distorts into two arc segments meeting at the
monopole worldlines (cf.~Fig.~1 of Ref.~\cite{Tranchedone:2026lav}). The
worldsheet bounce action is then~\cite{Tranchedone:2026lav}
\begin{equation}
  S_B(T) \;=\; 2\kappa\!\left[
    \arcsin\left(\frac{T_0}{T}\right)
    +\frac{T_0}{T}\sqrt{1-\!\Bigl(\frac{T_0}{T}\Bigr)^{\!2}}
  \right].
  \label{eq:SB_T_mech}
\end{equation}
The expression on the right-hand side is valid in the interval $T_0 \le T<\infty$.
Two limits provide useful cross-checks. As the lower limit, $T\to T_0$,
$\arcsin(T_0/T)\to\pi/2$ and the square root vanishes, so
$S_B(T)\to\pi\kappa = S_B(0)$,  matching the zero-temperature
expression of Eq.~\eqref{eq:Gamma0_mech}. For even lower temperatures $0\le T < T_0$, the bounce action is frozen at this constant value $\pi\kappa$. In the opposite, high-temperature limit,
$T\gg T_0$, Eq.~\eqref{eq:SB_T_mech} reduces to
\begin{equation}
  S_B(T)\;\xrightarrow{\;T\gg T_0\;}\;\frac{2M_m}{T},
  \label{eq:SB_highT_mech}
\end{equation}
so the bounce grows only as $\sqrt{\kappa}$, since $M_m\propto\sqrt{\mu_{\rm str}\kappa}$.
Physically, the arc-like solution interpolates between pure quantum
tunnelling and pure thermal hopping over the worldsheet barrier: in the
high-$T$ regime the dominant contribution is sub-critical thermal nucleation
of true-vacuum bubbles, which subsequently expand to infinity.

\subsection{Network history: early breaking, super-horizon segments,
late destruction}

We assume the dark sector is in internal thermal equilibrium, and that the strings form at the dark electroweak transition at the nucleation
temperature $T_{\rm nuc}$, computed from the one-loop finite-temperature effective potential detailed in Section~\ref{sec:thermal}. \rev{The string network is then destroyed at a much lower temperature $T_d$, which we estimate below.}
\rev{The temperature $\Tnuc$ fixes only the beginning of this history, and further scales appear between formation and destruction, none of which we compute. First, bubbles must collide and percolate before the network exists: percolation completes at a temperature $T_{\rm perc}$ somewhat below $\Tnuc$, and the collisions can reheat the plasma to $T_{\rm rh}$, so the temperature the newly formed strings actually experience lies in a narrow corridor around $\Tnuc$ rather than exactly at it. Because the fragmentation rate is exponentially sensitive to temperature, the width of this corridor feeds directly into the uncertainty of the bounce-action target quantified below. Determining $T_{\rm perc}$ and $T_{\rm rh}$ is a well-developed but separate computation, so far carried out for dark U(1) and scalar sectors~\cite{Ertas:2021xeh,Bringmann:2023opz,Athron:2023xlk}; to our knowledge no such calculation exists for a dark $\mathrm{SU}(2)\times\mathrm{U}(1)$ sector, and we leave it to future work. Second, the newly formed strings are immersed in the dark plasma, which exerts friction on their motion: in the near-semi-local regime $g\ll g'$ the charged $W^\pm$ are light at formation, $M_W=g\eta/2\ll\Tnuc$, and their scattering off the string cores damps string motion until the bath dilutes, at $T_{\rm fric}\sim\mstr/\MPl=\Gmu\MPl\sim10^{12}\,$GeV. Friction does not enter the local worldsheet action $\SB(T)$, and the super-horizon segments that set the cutoff are simply stretched by the expansion, so friction is neglected in our cutoff estimate; friction can, however, modify network relaxation and loop production, so the predicted amplitude and detailed spectral shape carry a network-evolution uncertainty that we do not model. Our use of $\Tnuc$ as the formation and fragmentation scale, and the comoving stretching assumed below, should therefore be read as explicit working assumptions rather than as a precision treatment of the full network history.}

\begin{figure}[t!]
    \centering \includegraphics[width=\linewidth]{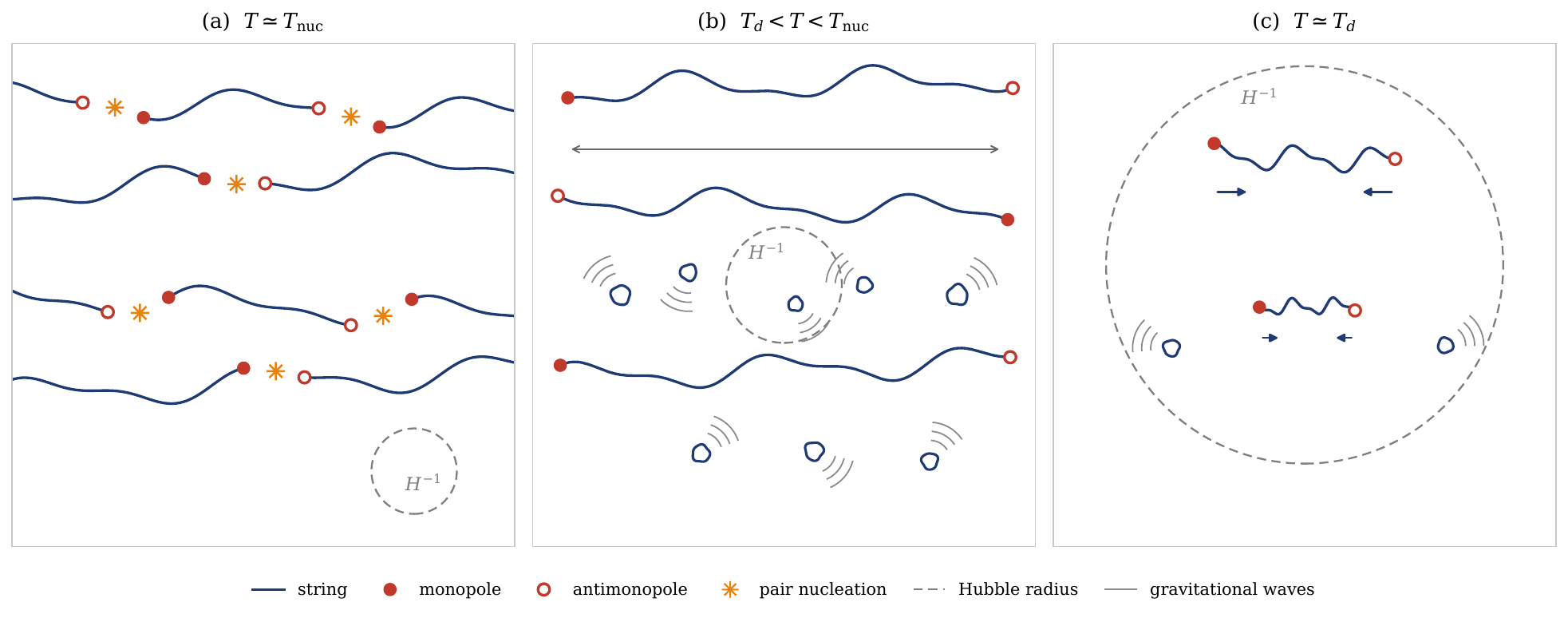}
    \caption{\rev{Illustration of the finite-temperature decay mechanism of the
    metastable string network. In all panels, filled and open circles denote
    monopoles and antimonopoles respectively, orange asterisks mark thermal
    monopole--antimonopole pair nucleation, the dashed circle indicates the Hubble
    radius, and grey arcs denote gravitational wave emission.
    (a) At formation, $T\simeq T_{\rm nuc}$, thermal monopole--antimonopole
    nucleation severs the long strings. Each break
    terminates one of the two new string ends on a monopole and the other on an
    antimonopole. The resulting
    segments remain much larger than the Hubble radius, and the long strings
    continue beyond the boundary of the panel.
    (b) For $T_d<T<T_{\rm nuc}$ thermal breaking has frozen out: the comoving
    endpoint separation is fixed while the physical separation stretches with the
    scale factor, so the endpoints remain super-horizon, as indicated by the
    double-headed arrow. The gravitational wave signal is emitted by the small
    closed loops chopped off the long strings by intercommutation. These loops are
    sub-horizon and carry no endpoints. We note that their radii are below one per cent of the Hubble radius and the loop sizes are exaggerated here for visibility.
    (c) At $T\simeq T_d$ the endpoints of a segment come into causal contact, are
    pulled together by the string tension (arrows) and annihilate, terminating the
    scaling regime. Whether the collapsing segments oscillate, and thereby radiate
    gravitational waves in their own right, is debated:
    Ref.~\cite{Chitose:2025qyt} argues that friction against thermal fluctuations
    on the string suppresses such oscillations, rendering the segment contribution
    negligible. We do not rely on segment oscillation: throughout this work the
    signal originates from the loops of panel~(b).}}
    \label{fig:thermal-decay}
\end{figure}

The physical mechanism is summarised in Fig.~\ref{fig:thermal-decay}.
At formation, $T\simeq T_{\rm nuc}$ (left panel),
the Kibble mechanism produces a network of \rev{long strings}. At this temperature, thermal
monopole--antimonopole pair nucleation on the worldsheet is highly
efficient and rapidly chops these long strings into finite segments,
which nonetheless remain super-horizon. Once the temperature drops
well below $T_{\rm nuc}$, the worldsheet action $S_B(T)$ rises sharply
back toward its zero-temperature value $\pi\kappa$. The thermal breaking channel then effectively shuts off, and the segment population is
frozen. The network enters a quasi-scaling regime
(central panel of Fig.~\ref{fig:thermal-decay}): long strings interact, intercommute, and chop off sub-horizon loops in the usual way, and these loops
oscillate and emit gravitational waves over many Hubble times,
sourcing the standard scaling-regime $\Omega_{\rm gw}$ plateau exactly
as in the stable-string case. Meanwhile, the average segment length
$\bar{\ell}(T)$, fixed in comoving coordinates, is stretched by the
scale factor and eventually re-enters the Hubble horizon at the
destruction temperature $T_d$ (right panel of Fig.~\ref{fig:thermal-decay}), at
which the monopole--antimonopole pairs at the segment ends come into
causal contact and annihilate, terminating the scaling regime. 
\rev{For reference, the viable points considered below have a string-formation temperature $T_{\rm nuc}\simeq(4$--$6)\times10^{14}$~GeV, while the PTA-motivated network-destruction temperature is $T_d\sim 20$~keV.}
\rev{A formation temperature of this magnitude is an important cosmological
caveat. The thermal scenario requires post-inflationary reheating to populate the
dark sector to at least $T_{\rm nuc}$ and requires sufficient thermal contact
within the dark sector. For comparison, the current tensor bound implies an
instantaneous-reheating upper limit of order
$T_{\rm rh}^{\rm max}\simeq6\times10^{15}$~GeV~\cite{BICEP:2021xfz}. }
The characteristic length $\bar{\ell}(T)$ follows from a Boltzmann equation for the segment
number density and depends strongly on the temperature profile of $\Gamma(T)$, the probability per unit time and unit string length for a monopole--antimonopole pair to nucleate on the worldsheet and break the string, defined in Eq.~\eqref{eq:Gamma}. Its temperature dependence enters through the compactification of the Euclidean worldsheet time on a circle of size $1/T$, encoded in Eq.~\eqref{eq:SB_T_mech}. The defect parameters $\mu_{\rm str}$ and $M_m$ are evaluated at their zero-temperature values; their finite-temperature counterparts require a dedicated computation of the thermal defect profiles and free energies, which we leave to future work. We collect the derivation of $\bar\ell(T)$ in Appendix~\ref{app:string-length}.
For the discussion that follows, we neglect re-percolation effects (the recombination of disconnected segments) and loop-formation effects, and comment on these below.

In the zero-temperature limit, the decay rate is constant, $\Gamma(T)=\Gamma_0$. As a result, as the
Universe cools, the average segment length decreases in proportion to the Hubble scale (see Appendix~\ref{app:string-length})
\begin{equation}
\label{eq:ell_avg}
    \bar{\ell}(T) \sim \frac{H(T)}{\Gamma_0}\,.
\end{equation}
The network decay temperature $T_d$ can then be obtained from
\begin{equation}
    \label{eq:gamma_T0}
    \Gamma_0\sim H_d^2\,,
\end{equation}
where we defined $H_d=H(T_d)$. 
This is the estimate typically used in the literature and was employed for the same model in Ref.~\cite{Ingoldby:2025wcl}.
Indeed, substituting the estimate for $\Gamma_0$ in~\eqref{eq:gamma_T0} into 
Eq.~\eqref{eq:ell_avg} gives the expected relation
$\bar{\ell}(T_d)\sim H_d^{-1}$
for the network destruction to occur.

As shown in Ref.~\cite{Tranchedone:2026lav}, thermal effects substantially modify this picture. The worldsheet action $S_B(T)$ falls steeply with temperature in the range $T_0<T\le T_\nuc$, so the breaking rate $\Gamma(T)=(\mu_{\rm str}/2\pi)\,e^{-S_B(T)}$ rises steeply with $T$ and is largest at the formation temperature $T_\nuc$, making string breaking extremely efficient at the onset of the thermal phase. Once the temperature drops appreciably below $T_\nuc$, $S_B(T)$ grows rapidly back towards its zero-temperature value $\pi\kappa$ and the thermal breaking channel effectively shuts off. The brief, early-time burst of breaking events imprints a characteristic average length on the network (see Appendix~\ref{app:string-length})
\begin{equation}
\label{eq:ell_star}
    \bar{\ell}(T)\;\sim\;\frac{H_\nuc}{\Gamma_\nuc}\,,
\end{equation}
approximately independent of temperature, where
$H_\nuc \equiv H(T_\nuc) \simeq 1.66\sqrt{g_*}\,T_\nuc^2/M_{\rm Pl}$.
For the parameter range relevant to PTAs, the segments are still super-horizon at $T_\nuc$.
Then, the decay temperature can be obtained from
\begin{equation}
\label{eq:gamma_Tnuc}
    \Gamma_\nuc \sim H_\nuc H_d\,.
\end{equation}
This differs substantially from the zero-temperature result of Eq.~\eqref{eq:gamma_T0}.

An extra subtlety has to be included. As noted in Ref.~\cite{Tranchedone:2026lav}, the relevant length scale for network survival is the endpoint separation, not the total contour length of the string. Since the string need not remain straight, the effective endpoint separation $\bar{\ell}_{\rm eff}(T)$ is on average smaller than $\bar{\ell}(T)$, $\bar{\ell}_{\rm eff}(T)\ll \bar{\ell}(T)$.
A random-walk treatment of the network at formation
gives~\cite{Tranchedone:2026lav}
\begin{equation}
  \bar{\ell}_{\rm eff}(T_\nuc) \;\simeq\; \sqrt{\frac{\bar{\ell}_\nuc}{H_\nuc}}=\Gamma_\nuc^{-1/2}\,,
  \label{eq:ell0_mech}
\end{equation}
where we have assumed that at $T_\nuc$ the strings are locally straight
on sub-Hubble scales. Relaxing this assumption to include random-walk
effects at formation would further reduce $\bar{\ell}_{\rm eff}$ and hence \rev{raise} $T_d$, so that Eq.~\eqref{eq:ell0_mech} is, if anything,
a conservative upper bound on the segment lifetime; the PTA target
$S_B(T_\nuc)\simeq \rev{109}$ derived below should be read as a \rev{lower} bound on the required bounce action. We leave a detailed treatment to
future work and take Eq.~\eqref{eq:ell0_mech} as our working assumption. As the Universe expands the endpoint separation is stretched by the
scale factor, $a(T)/a(T_\nuc)=T_\nuc/T$, so the condition of
Eq.~\eqref{eq:gamma_Tnuc} becomes
\begin{equation}
    H_d^{-1}\;\sim\;\bar{\ell}_{\rm eff}(T_d)
                  \;=\;\bar{\ell}_{\rm eff}(T_\nuc)\,\frac{T_\nuc}{T_d}
                  \;=\;\Gamma_\nuc^{-1/2}\,\frac{T_\nuc}{T_d}\,.
\end{equation}
Following Ref.~\cite{Tranchedone:2026lav}, we use the standard
convention $H(T)\sim T^2/\MPl$ for the network-destruction estimate,
absorbing the $\mathcal{O}(g_*^{1/2})$ Hubble prefactor into
the overall calibration. The decay temperature is then
\begin{equation}
    T_d \;\simeq\; \MPl\,
                \sqrt{\frac{\mstr}{2\pi\,T_\nuc^2}}\;
                e^{-S_B(T_\nuc)/2}
        \;=\; \frac{\MPl\,\eta\sqrt{\alpha_{\rm str}/(2\pi)}}{T_\nuc}\,
                e^{-S_B(T_\nuc)/2}\,.
    \label{eq:Td_mech}
\end{equation}
It is useful to compare Eq.~\eqref{eq:Td_mech} with the destruction
temperature in the zero-temperature picture, since PTA data directly
constrain $T_d$ (via the low-frequency cutoff \rev{$\fstar$}) but not the
bounce action itself. The network survives the scaling regime until the Universe cools to a temperature $T_d$, at which point the Hubble expansion rate becomes comparable to the constant tunnelling rate of Eq.~\eqref{eq:Gamma0_mech}, $H^2(T_d) \sim \Gamma_0$. Using $H\sim T^2/M_{\rm Pl}$, this yields
\begin{equation}
  T_d^{(0)} \;\sim\; M_{\rm Pl}(G_N\mu_{\rm str})^{1/4}\,e^{-\pi\kappa/4}\,,
  \label{eq:Td_zeroT_mech}
\end{equation}
where $G_N$ is Newton's constant.
The crucial difference between Eqs.~\eqref{eq:Td_mech}
and~\eqref{eq:Td_zeroT_mech} is the factor in the exponent: the
zero-temperature destruction temperature is suppressed as
$e^{-S_B(0)/4}$, whereas the finite-temperature one is suppressed as
$e^{-S_B(T_{\rm nuc})/2}$. The factor of four in the zero temperature case arises
because $T_d$ enters as a fourth power through $H^2\propto T^4$ while the
factor of two in the finite-$T$ case arises from the square root in the
random-walk segment length of Eq.~\eqref{eq:ell0_mech}. The two exponent factors
are direct consequences of two genuinely different network-destruction
histories. In the zero temperature case the network must wait for Hubble to shrink
to match a constant tunnelling rate. In the finite-$T$ case the network's
lifetime is set by a geometric scale fixed at formation. They are not
re-parametrisations of the same condition.

\rev{We relate the destruction temperature to the gravitational wave spectrum
using Ref.~\cite{Asl:2026zpj}, which describe the spectrum of a
metastable network in terms of three parameters: the string tension
$G_N\mu_{\rm str}$, the time scale $t_{\rm LB}$ of loop breaking by monopole
nucleation, and the time scale $t_{\rm NC}$ of network collapse, at which
string segments with monopoles on their ends re-enter the Hubble horizon. In
our scenario these two time scales are widely separated. The burst of thermally
induced breaking at formation fixes the comoving endpoint separation of the
segments, after which the thermal channel shuts off. The residual
zero-temperature nucleation rate is negligible across the entire viable region,
where $S_B(0)=\pi\kappa\gtrsim 700$. Closed sub-Hubble loops are therefore
unaffected by monopole nucleation over the age of the Universe,
$t_{\rm LB}\gg t_0$ in the notation of Ref.~\cite{Asl:2026zpj}, while the
long-string network disappears at the network-collapse time
$t_{\rm NC}=t(T_d)$, with $T_d$ given by Eq.~\eqref{eq:Td_mech}. The
gravitational wave spectrum therefore falls into the quasi-stable class of
Refs.~\cite{Lazarides:2022jgr,Lazarides:2023ksx,Asl:2026zpj}, and the relevant
low-frequency cutoff is not the loop-breaking frequency $\flow$ of
Refs.~\cite{Buchmuller:2021mbb,Tranchedone:2026lav}, which characterises the
depletion of sub-Hubble loops by monopole nucleation, but the network-collapse
frequency. In this class the spectrum falls off approximately as $f^{3}$
below the cutoff, rather than as the $f^{2}$ of the tuned case
$t_{\rm LB}\simeq t_{\rm NC}$; since our analysis uses only the position of
the cutoff, this affects its interpretation rather than the parameter map.
Across the strip the hierarchy $t_{\rm NC}\ll t_{\rm end}\ll t_{\rm LB}$ holds
with enormous margin, where $t_{\rm end}$ is the time at which the last loops
finish radiating, so loop breaking is negligible throughout the emission
history. The cutoff is (see e.g. ~Eq.~(48) of Ref.~\cite{Asl:2026zpj})
\begin{equation}
\fstar 
% \;=\; \frac{f_{\rm NC}}{\left(1+\chi_r\right)^{1/2}}
\;\approx\;
\left(\frac{8H_r^0}{\alpha_r\,\Gamma_g\,G_N\mu_{\rm str}\,t_{\rm NC}}\right)^{\!1/2}
% \qquad
% f_{\rm NC}\;=\;\frac{2}{\Gamma_g\, G_N\mu_{\rm str}}
% \left(\frac{2H_r^0}{t_{\rm NC}}\right)^{\!1/2}
\,,
\label{eq:fstar_mech}
\end{equation}
where 
% $\chi_r=\alpha_r/(\Gamma_g G_N\mu_{\rm str})\gg1$, 
$\alpha_r\simeq0.1$ is
the loop-size parameter, $\Gamma_g\simeq 50$ is the gravitational radiation
efficiency of loops, and $H_r^0=H_0\,\Omega_r^{1/2}$ is the radiation-scaled
Hubble constant, with $h^2\Omega_r\simeq4.18\times10^{-5}$. To express $\fstar$
in terms of the decay temperature we use the radiation-domination relation
\begin{equation}
t(T) \simeq 0.30\, g_*^{-1/2}\,\frac{M_{\rm Pl}}{T^2}\,,
\label{eq:radiation_time_temperature}
\end{equation}
so that $\fstar$, like the loop-breaking cutoff it replaces, is linear in $T_d$
at fixed $G_N\mu_{\rm str}$,
\begin{equation}
\fstar \;\simeq\; 5.1\times10^{-4}
\left(\frac{0.1}{\alpha_r}\right)^{\!1/2}
\left(\frac{50}{\Gamma_g}\right)^{\!1/2}
\left(\frac{10^{-7}}{G_N\mu_{\rm str}}\right)^{\!1/2}
\left(\frac{g_*(T_d)}{3.36}\right)^{\!1/4}
\left( \frac{T_d}{\rm GeV}\right)\;{\rm Hz}\,.
\label{eq:fstar_Td}
\end{equation}
Setting $\fstar=10^{-8}~{\rm Hz}$ at $G_N\mu_{\rm str}=10^{-7}$,
$\alpha_r=0.1$ and $\Gamma_g=50$ then gives a target destruction temperature
\begin{equation}
T_d^{\rm target}\simeq 20~{\rm keV}\,,
\end{equation}
corresponding to a network-collapse time $t_{\rm NC}\simeq 3\times10^{3}$~s. We
take the number of relativistic degrees of freedom to be $g_*=3.36$,
appropriate for temperatures below the electron mass, where the radiation bath
consists of photons and decoupled neutrinos. A string network collapsing on
this timescale is \rev{of the order relevant to the PTA fits of Ref.}~\cite{Asl:2026zpj}.
Inverting
Eq.~\eqref{eq:Td_mech} then defines the central PTA condition used
throughout this work,
\begin{equation}
    S_B(T_{\rm nuc}) \;\simeq\;
    2\ln\!\left(\frac{M_{\rm Pl}\,\eta\sqrt{\alpha_{\rm str}/(2\pi)}}
                     {T_{\rm nuc}\,T_d^{\rm target}}\right)
    \;\simeq\;109\,.
  \label{eq:SBtarget_mech}
\end{equation}
For the parameter range considered here ($\eta\sim 10^{15}$~GeV,
$T_{\rm nuc}/\eta\simeq 0.4$--$0.6$, $\alpha_{\rm str}\sim\mathcal{O}(1)$), the
argument of the logarithm is $\sim 5\times10^{23}$, giving the quoted central
value. The remaining $\mathcal{O}(g_*^{1/2})$ Hubble prefactor shifts this
estimate by $\Delta S_B\lesssim 7$. Across the viable first-order points, the ratio of critical to nucleation temperature is $T_\star/T_{\rm nuc}\simeq1.03$ on average and never exceeds $1.10$, indicating modest supercooling, though this does not replace a calculation of percolation and reheating; and since, within our fixed-defect approximation, $S_B\simeq2M_m/T$, even a few-per-cent shift in the effective fragmentation temperature moves $S_B$ by several units, comparable to the prefactor shift above. Together with uncertainties in the mapping
between $T_{\rm nuc}$ and the actual fragmentation temperature, this means that
$S_B(T_{\rm nuc})\simeq 109$ is a phenomenological central target rather than a
precision prediction.}

%%%==========================================================================
\section{Thermal phase transition and nucleation temperature}
\label{sec:thermal}
%%%==========================================================================
\rev{The string network forms during the dark-electroweak transition, in which the Higgs field $\Phi$ acquires its vacuum expectation value; the network's properties are set by the thermodynamic history of this transition. A first-order transition is not a prerequisite for string formation: the
Kibble mechanism requires only that the order parameter be uncorrelated between causally disconnected regions, and
strings form equally well in a continuous transition. What a first-order
transition does provide is an approximate \emph{formation temperature}. When
bubbles nucleate and collide, the phase of the order parameter is uncorrelated
between neighbouring bubbles, so the initial correlation length is set by the
bubble size at percolation and the network forms within a narrow interval around
$\Tnuc$. For a continuous transition, by contrast, the formation temperature is
fixed by a Kibble--Zurek freeze-out condition \cite{KibbleZurek,zurek1985cosmological}, in which the relaxation rate of the order parameter is compared with the quench rate $\sim H$. Evaluating it
requires the dissipative dynamics of $\phi$ in the dark plasma, which we do not
compute here. We therefore quote $\Tnuc$ as the formation temperature where the
transition proceeds by bubble nucleation, and our quantitative parameter scan is
accordingly restricted to points for which a nucleation temperature can be
determined; we do not interpret the absence of bubble nucleation as an absence
of strings, but assigning a formation scale to a continuous transition would
require the Kibble--Zurek treatment described above.
Separately, in a first-order transition the interplay of monopoles,
string segments and expanding bubble walls can admit additional configurations,
such as the slingshot mechanism of Ref.~\cite{Bachmaier:2026bgw}, in which a
source crossing the wall from the unconfined into the Higgsed phase is left
tethered to it by a stretched flux tube. In the one-scale model considered here
the vacuum manifold is $S^3$, with $\pi_2=0$, so no topological monopoles exist
in the symmetric phase to be swept up by the walls; such configurations are
therefore expected to be considerably less relevant than in two-scale
constructions, where monopoles are produced at an earlier transition.}

 \rev{To determine $T_{\rm nuc}$, we} work with the finite-temperature effective potential along
the neutral Higgs direction,
\begin{equation}
  \Phi(x) \;=\; \frac{1}{\sqrt{2}}
              \begin{pmatrix} 0 \\ \phi(x) \end{pmatrix},
  \label{eq:phi_param}
\end{equation}
so that the zero-temperature vacuum lies at $\phi=\eta$. We decompose the
effective potential as
\begin{equation}
  V_{\rm eff}(\phi,T)
  \;=\;
  V_{\rm tree}(\phi)
  +V_{\rm CW}(\phi)
  +V_{\rm ct}(\phi)
  +V_T(\phi,T)
  +V_{\rm daisy}(\phi,T)\,,
  \label{eq:Veff_full}
\end{equation}
where $V_{\rm CW}$ is the one-loop Coleman--Weinberg
correction~\cite{Coleman:1973jx}, $V_{\rm ct}$ is a finite counterterm
chosen to preserve the tree-level vacuum and Higgs mass at $\phi=\eta$,
$V_T$ is the one-loop bosonic thermal contribution
\cite{Dolan:1973qd,Quiros:1999jp} and $V_{\rm daisy}$ implements daisy
(ring) resummation of the longitudinal gauge-boson and scalar zero modes~\cite{Arnold:1992rz}. We adopt the prescription used in Ref.~\cite{Arnold:1992rz}, in which only the bosonic Matsubara zero modes are resummed. This is reliable along the bounce trajectory whenever the relevant masses satisfy $m_i(\phi)\lesssim T$.  The explicit
field-dependent masses, counterterm conditions, thermal integrals and Debye self-energies entering Eq.~\eqref{eq:Veff_full} are collected in
Appendix~\ref{app:thermal_potential}. Following standard practice
we shift the potential at every temperature so that $V_{\rm eff}(0,T)=0$.

\begin{figure}[t!]
    \centering
    \includegraphics[width=\linewidth]{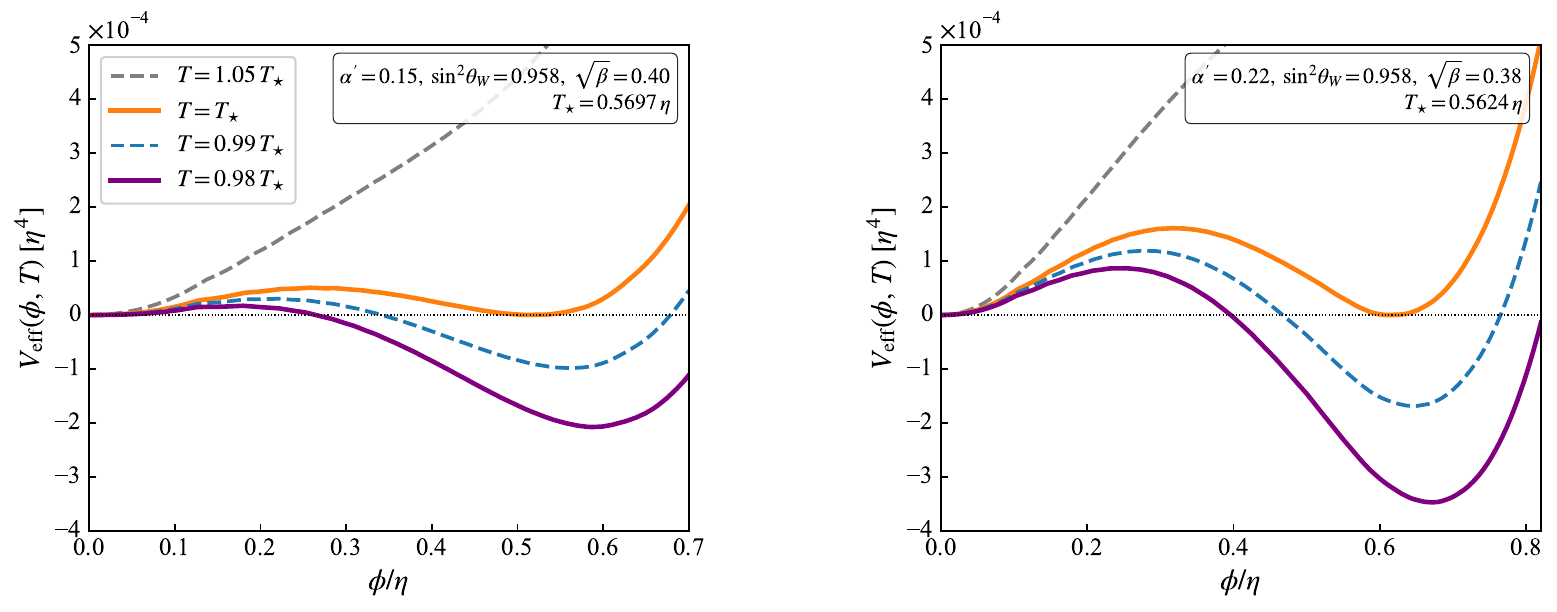}
    \caption{Finite-temperature effective potential $V_\text{eff}(\phi, T)$ as defined in Eq.~\eqref{eq:Veff_full}, evaluated at four characteristic temperatures for two PTA-compatible benchmarks in the dark $\mathrm{SU}(2)\times\mathrm{U}(1)$ sector: $(\alpha', \sin^2\theta_w, \sqrt{\beta}) = (0.15,\, 0.958,\, 0.40)$ (left) and $(0.22,\, 0.958,\, 0.38)$ (right). The two minima become degenerate at $T=T_\star$. Both panels share a common $y$-axis; the right-hand plot has a barrier roughly four times taller than the left, reflecting a more strongly first-order transition.}
    \label{fig:potentials}
\end{figure}
Figure~\ref{fig:potentials} shows the evolution of the potential with temperature for two representative benchmarks. At high temperature, the symmetric phase is favoured; as the Universe
cools, a second minimum develops, and the critical temperature $T_\star$ is defined by the degeneracy of the symmetric and broken minima. Below $T_\star$, the broken minimum is the true vacuum. The string network forms when bubbles of the broken phase nucleate and percolate. Throughout our work we use the nucleation temperature $\Tnuc<T_\star$ as a reference scale for this. As shown in Fig.~\ref{fig:potentials}, the choice of model parameters changes the height of the potential barrier between the two minima, thus appreciably affecting the value of $\Tnuc$. We compute $\Tnuc$ from the three-dimensional Euclidean action $S_3(T)$, which is conceptually distinct from the worldsheet string-breaking action $\SB(T)$ of Section~\ref{sec:mechanism}. $S_3$ controls the nucleation of bubbles of the broken phase that create the string network, while $\SB$ controls the subsequent nucleation of monopole--antimonopole pairs on the worldsheets of the strings that result.

The $O(3)$-symmetric critical-bubble profile obeys
\begin{equation}
  \frac{d^2\phi}{dr^2}
  + \frac{2}{r}\frac{d\phi}{dr}
  = \frac{\partial V_{\rm eff}(\phi,T)}{\partial\phi}\,,
  \qquad
  \left.\frac{d\phi}{dr}\right|_{r=0}=0\,,
  \qquad
  \phi(r\to\infty)=0\,,
  \label{eq:S3_bounce_eq}
\end{equation}
and the corresponding Euclidean action is
\begin{equation}
  S_3(T)
  \;=\;
  4\pi\int_0^\infty dr\,r^2
  \left[\frac12\!\left(\frac{d\phi}{dr}\right)^{\!2}
        + V_{\rm eff}(\phi,T)\right]\,.
  \label{eq:S3_action}
\end{equation}
The nucleation temperature is defined by requiring
approximately one bubble per Hubble four-volume,
$\Gamma_{\rm PT}(\Tnuc)\,H^{-4}(\Tnuc)\sim 1$, with the phase-transition (PT) rate and Hubble rate given by
\begin{equation}
  \Gamma_{\rm PT}(T)\;\simeq\;T^4\,e^{-S_3(T)/T}\,,
  \qquad
  H(T)\;=\;1.66\sqrt{g_*}\,\frac{T^2}{\MPl}\,,
  \label{eq:PT_rate}
\end{equation}
where $g_*$ counts relativistic degrees of freedom in the radiation bath at
the transition. Inverting Eq.~\eqref{eq:PT_rate} gives the
nucleation condition
\begin{equation}
  \frac{S_3(\Tnuc)}{\Tnuc}
\simeq 
  4\ln\!\left[\frac{\MPl}{1.66\sqrt{g_*}\,\Tnuc}\right]\,.
  \label{eq:S3criterion}
\end{equation}
The numerical value of the threshold $S_3/\Tnuc$ depends on the scale of
the transition through the logarithm in Eq.~\eqref{eq:S3criterion}. For the
$\Gmu=10^{-7}$ benchmark used in this work, the string-forming scale is
$\eta\sim10^{15}\,{\rm GeV}$, the nucleation temperature is
\rev{$\Tnuc\simeq(0.4$--$0.6)\,\eta$} across the viable strip, and $g_*\simeq100$
at the GUT-scale transition. Eq.~\eqref{eq:S3criterion} then gives
\begin{equation}
  \frac{S_3(\Tnuc)}{\Tnuc} \;\simeq\; 29\text{--}31\,.
  \label{eq:S3criterion_GUT}
\end{equation}
% This is to be contrasted with the much higher value of the electroweak-scale rule-of-thumb criterion $S_3/T\simeq140$~\cite{Dine:1992wr}.
% \rev{We note that the dark sector itself contributes only $g_{\rm dark}=12$ degrees of
% freedom --- four gauge bosons and a complex doublet --- so $g_*$ at the
% transition is dominated by the visible sector.} \rev{We further note that the truncated, gauge-fixed (Landau-gauge) effective potential carries a residual gauge dependence, so the critical and nucleation temperatures extracted from it are subject to the usual perturbative ambiguities~\cite{Nielsen:1975fs,Patel:2011th}.}
% 
% For each parameter-space point $(\ap,\sw,\sqrt{\beta})$ we construct $V_{\rm eff}(\phi,T)$ from Eq.~\eqref{eq:Veff_full}, identify the symmetry breaking vacuum and the critical temperature $T_\star$, and integrate the
% O(3)-symmetric bounce Eq.~\eqref{eq:S3_bounce_eq} using
% \texttt{CosmoTransitions}~\cite{Wainwright:2011kj} to determine the nucleation temperature and the vacuum expectation value.

This is to be contrasted with the much higher value of the electroweak-scale rule-of-thumb criterion $S_3/T\simeq140$~\cite{Dine:1992wr}.
\rev{We note that the dark sector itself contributes only $g_{\rm dark}=12$ degrees of
freedom --- four gauge bosons and a complex doublet --- so $g_*$ at the
transition is dominated by the visible sector.}
For each parameter-space point $(\ap,\sw,\sqrt{\beta})$ we construct $V_{\rm eff}(\phi,T)$ from Eq.~\eqref{eq:Veff_full}, identify the symmetry breaking vacuum and the critical temperature $T_\star$, and integrate the
O(3)-symmetric bounce Eq.~\eqref{eq:S3_bounce_eq} using
\texttt{CosmoTransitions}~\cite{Wainwright:2011kj} to determine the nucleation temperature and the vacuum expectation value.
\rev{Where the transition is continuous, or so weak that the nucleation
condition of Eq.~\eqref{eq:S3criterion} is never met, no $\Tnuc$ exists and the
point is excluded from the quantitative scan, as discussed in
Section~\ref{sec:thermal}.}

%%%==========================================================
\section{Results and Discussion}
\label{sec:results}
%%%==========================================================
\begin{figure}[t]
\centering
\includegraphics[width=\textwidth]{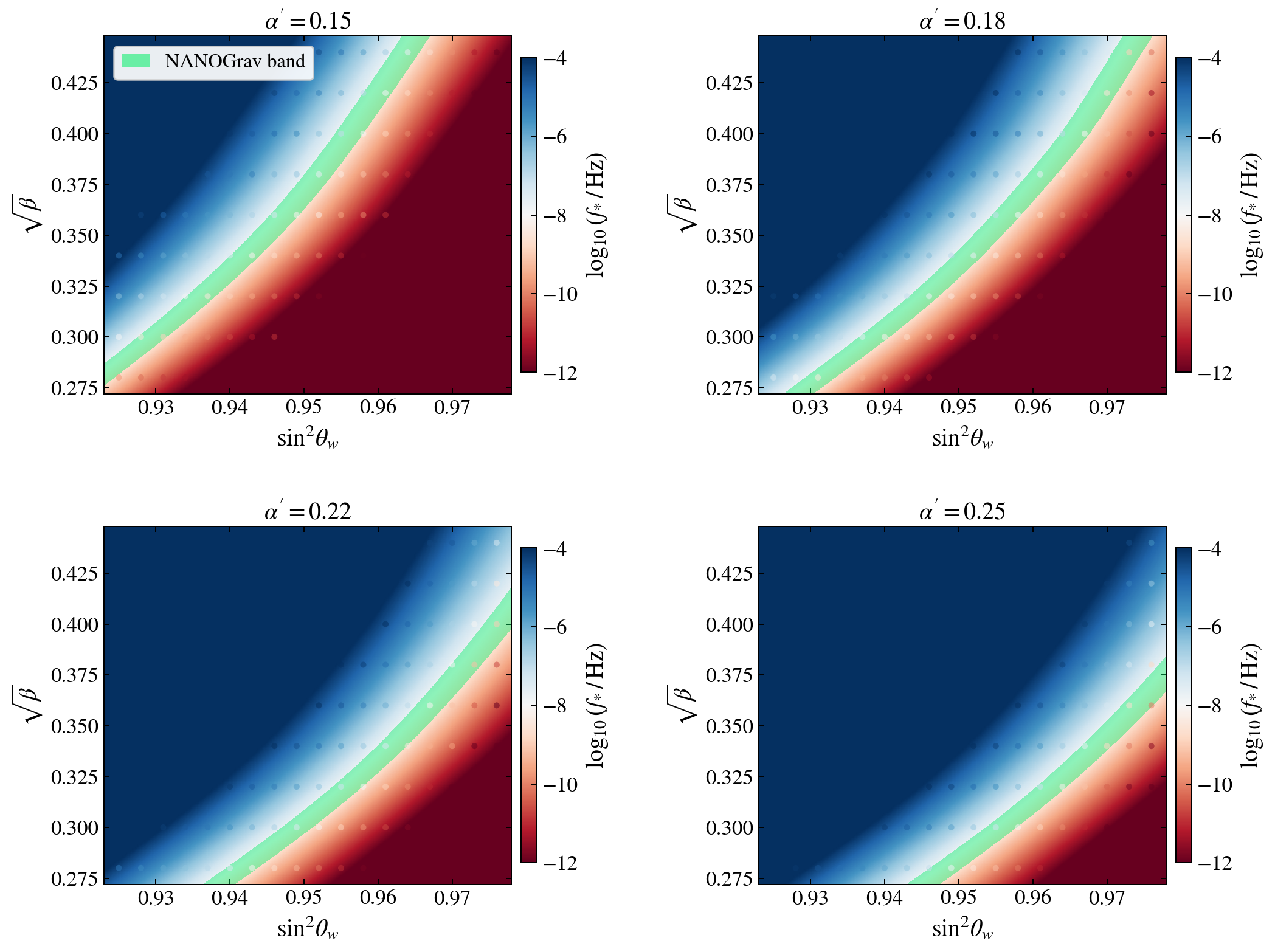}
\caption{\rev{$\log_{10}(\fstar/\text{Hz})$} across the PTA strip for four
values of $\ap$. Green region: approximate PTA-motivated 2--20\,nHz
band; the corresponding worldsheet action is typically
\rev{$\SB(\Tnuc)\sim 108$--$113$} for the benchmark $G_N\mu_{\rm str}=10^{-7}$
and $\Gamma_g=50$, with mild residual dependence on $\Tnuc/\eta$ and
$\alpha_{\rm str}$. We note that the axes range of the plots is within the zero-temperature classical stability region \cite{James:1992wb}.}
\label{fig:strip}
\end{figure}
We identify the region of model parameter space in which the
finite-temperature destruction of the metastable string network produces a
low-frequency cutoff in the PTA band. For this purpose we use the interval
\begin{equation}
  2\times 10^{-9}\,{\rm Hz}
  \;\lesssim\;
  \rev{\fstar}
  \;\lesssim\;
  2\times 10^{-8}\,{\rm Hz}\,,
  \label{eq:pta_window}
\end{equation}
as a representative target range. \rev{Through Eqs.~\eqref{eq:Td_mech},
\eqref{eq:fstar_mech} and~\eqref{eq:fstar_Td}, the edges of this band map onto the
finite-temperature worldsheet bounce-action window
\begin{equation}
  % 108.3
  % \;\lesssim\;
  % \SB(\Tnuc)
  % \;\lesssim\;
  % 112.9\,,
    108
  \;\lesssim\;
  \SB(\Tnuc)
  \;\lesssim\;
  113\,,
  \label{eq:SB_window}
\end{equation}
evaluated at the reference point $(\ap,\sw,\sqrt{\beta})=(0.15,0.958,0.40)$, in
good agreement with the realised range 
%$\SB(\Tnuc)\simeq108.6$--$112.7$ 
$\SB(\Tnuc)\simeq108$--$113$ of the
in-band scan points; $\SB(\Tnuc)\simeq109.6$ at $\fstar=10^{-8}$~Hz} for the benchmark
$G_N\mu_{\rm str}=10^{-7}$ and $\Gamma_g=50$. The precise mapping
between $\SB(\Tnuc)$ and \rev{$\fstar$} at each point also depends mildly on
$\Tnuc/\eta$ and $\alpha_{\rm str}$ through Eq.~\eqref{eq:Td_mech}. Here $\Gamma_g$ denotes 
usual dimensionless gravitational radiation efficiency of a string loop,
defined by
$  P_{\rm GW}=\Gamma_g\,G_N\mu_{\rm str}^2$,
where we have used the subscript $g$ to distinguish this quantity from the worldsheet
breaking rate $\Gamma$ appearing in Eq.~\eqref{eq:Gamma}.
Figure~\ref{fig:strip} displays this condition across the
$(\sw,\sqrt{\beta})$ plane for representative values of the dark
fine-structure constant $\ap$. For each point we determine the nucleation
temperature $\Tnuc$ from the finite-temperature effective potential,
evaluate the worldsheet bounce action $\SB(\Tnuc)$, and convert the corresponding
network-destruction temperature into the low-frequency cutoff \rev{$\fstar$}. The
green band marks the PTA-motivated interval of Eq.~\eqref{eq:pta_window},
or equivalently the action window of Eq.~\eqref{eq:SB_window}.
The main feature of Fig.~\ref{fig:strip} is a continuous, approximately diagonal strip in the $(\sw,\sqrt{\beta})$ plane. Representative points close to the central contour span the ranges
% {\color{red} is the range $\SB(\Tnuc)\simeq 109\text{--}111$ or $\SB(\Tnuc)\simeq 109\text{--}113$?}
\begin{equation}
  \rev{\kappa\simeq 230\text{--}470\,,\qquad
  \Tnuc/\eta\simeq 0.4\text{--}0.6\,,\qquad
  \SB(\Tnuc)\simeq 108\text{--}113\,.}
\end{equation}
Thus, the finite-temperature PTA-compatible region lies at values of $\kappa$ of order several hundred, substantially above the zero-temperature window $\kappa\simeq 56$--$72$ identified in Ref.~\cite{Ingoldby:2025wcl}.
The origin of the diagonal strip can be understood analytically from the high-temperature limit of the worldsheet bounce. In this regime,
\begin{equation}
  \SB(\Tnuc)
  \;\simeq\;
  \frac{2M_m}{\Tnuc}
  =
  \frac{2\eta}{\Tnuc}\sqrt{\kappa\,\astr(\beta)}\,,
  \label{eq:SB_highT_scan}
\end{equation}
where we used the relation $\kappa = M_m^2/\mu_{\rm str} = M_m^2/(\alpha_{\rm str}\eta^2)$.
Using Eq.~\eqref{eq:kappa_ap_sw} we simplify the argument of the square root, observing that the leading dependence on the dimensionless string tension parameter cancels:
\begin{equation}
  \kappa\,\astr(\beta)
  =
  \left[
  \frac{2\pi\sw}
       {\ap(1-\sw)\astr(\beta)}
  \right]\astr(\beta)
  =
  \frac{2\pi\sw}{\ap(1-\sw)}\,,
  \label{eq:astr_cancel}
\end{equation}
and thus the leading high-temperature approximation becomes
\begin{equation}
  \SB(\Tnuc)
  \;\simeq\;
  \frac{2\eta}{\Tnuc}
  \sqrt{
    \frac{2\pi\,\sw}
         {\ap(1-\sw)}
  }\, .
  \label{eq:SB_strip_leading}
\end{equation}
With the explicit dependence on the string-tension function $\astr(\beta)$ cancelled between the endpoint mass and the definition of $\kappa$,
the remaining dependence on $\sqrt{\beta}$ enters through the thermal history, most importantly through $\Tnuc/\eta$, and through corrections to the strict high-temperature approximation. The diagonal strip in Fig.~\ref{fig:strip} is the locus on which these effects compensate so that $\SB(\Tnuc)$ remains close to the PTA target.
The $\ap$-dependence of the PTA-compatible strip follows directly from Eq.~\eqref{eq:SB_strip_leading}. At fixed $\Tnuc/\eta$, increasing $\ap$ lowers $\SB(\Tnuc)$, so maintaining $\SB(\Tnuc)\simeq\rev{109}$ requires a larger value of $\sw/(1-\sw)$, and hence larger $\sw$. The
remaining displacement in $\sqrt{\beta}$ reflects the correlated change in
the nucleation temperature across the thermal scan. This explains the
coherent upward shift in $\sw$ and the milder motion in $\sqrt{\beta}$ seen
as $\ap$ is varied in Fig.~\ref{fig:strip}.

It remains to check that the region selected by the PTA condition lies
in the part of the model where the embedded string configuration is classically
stable, so that it is well defined as a local minimum of the
classical energy functional.
At zero temperature the embedded $Z$-string can be a stable or
metastable local minimum of the energy functional in the near-semi-local
regime
\begin{equation}
  g \;\ll\; g'\,,
  \qquad
  \beta \;<\; 1\,,
  \label{eq:stab_T0}
\end{equation}
corresponding to $\sw\to1$ and $M_\Phi<M_Z$. The PTA-compatible points in
Fig.~\ref{fig:strip} lie in the range
\begin{equation}
  \sw\simeq0.92\text{--}0.97\,,
  \qquad
  \sqrt{\beta}\simeq0.28\text{--}0.45\,,
\end{equation}
or equivalently $\beta\simeq0.08\text{--}0.20$. They therefore lie
comfortably within the near-semi-local regime in which the embedded
string is known to be classically stable~\cite{James:1992wb}. \rev{The
parameters selected by the PTA condition thus lie, without any tuning,
inside the region where the embedded $Z$-string is a local minimum of
the energy functional.}
The stability criterion in Eq.~\eqref{eq:stab_T0} is, however, a
zero-temperature statement. Since the strings form at
\rev{$\Tnuc\simeq0.4$--$0.6\,\eta$}, a fully self-consistent treatment would
require analysing the stability of the embedded string using the
finite-temperature effective action at the transition. The corresponding
question has been studied in the Standard Model and in left--right
extensions in Ref.~\cite{Holman:1992rv}. There, it is shown that thermal effects can enlarge the stable region in the $(\sw,\beta)$ plane before the spinodal is reached,
although the Standard Model value $\sw\simeq0.23$ remains too small for
stability. For the present model this suggests that thermal corrections are unlikely
to invalidate the PTA-compatible strip, which already lies deep inside the
zero-temperature classical stability region. %Nevertheless, a dedicated finite-temperature stability analysis of the dark $\mathrm{SU}(2)\times\mathrm{U}(1)$ model would be required to determine the precise boundary. Such an analysis could also show whether the viable region extends to slightly smaller $\sw$ or larger $\sqrt{\beta}$ than the zero-temperature criterion would suggest. We leave this question to future work.

\section{Conclusions}
\label{sec:conclusions}

We have analysed metastable $Z$-strings in a one-scale dark-sector
$\mathrm{SU}(2)\times\mathrm{U}(1)$ gauge theory under the assumption that
the string-forming phase transition occurs in a thermal plasma.
Determining the nucleation temperature $\Tnuc$ from the one-loop
finite-temperature effective potential with daisy resummation, and
applying the finite-temperature worldsheet bounce analysis of
Ref.~\cite{Tranchedone:2026lav} at that temperature, we have mapped out
the region of microscopic parameter space in which the resulting
gravitational wave spectrum has its low-frequency cutoff in the PTA band.

The PTA-compatible region is a narrow, continuous diagonal strip in the
$(\sw,\sqrt{\beta})$ plane defined by $\SB(\Tnuc)\simeq \rev{109}$, present
across $\ap\in[0.12,0.25]$ (Fig.~\ref{fig:strip}). Representative
viable points have \rev{$\Tnuc/\eta\simeq 0.4$--$0.6$}, $\kIKT\sim$ a few
$\times 10^2$, and \rev{$\fstar\simeq 10^{-8}$~Hz}. The quantitative shift
relative to the cold-formation analysis of Ref.~\cite{Ingoldby:2025wcl}
is substantial: $\kIKT$ moves from $\approx 56$--$72$ to a few~$\times 10^2$,
and $\ap$ from $\approx 0.75$ to $\approx 0.10$--$0.25$. The origin is
purely the different parametric dependence of the relevant bounce action, $\SB(\Tnuc)\propto\sqrt{\kIKT}$ in the high-temperature regime versus $\SB(0)=\pi\kIKT$ at zero temperature; identifying the two thresholds would conflate two physically distinct destruction mechanisms.
The PTA-compatible strip lies comfortably within the near-semi-local
regime in which the embedded $Z$-string is expected to be metastable,
$\sw\to 1$ and $\beta<1$~\cite{Vachaspati:1991dz,Hindmarsh:1991jq,James:1992wb}:
\rev{the dynamically selected strip lies inside this classical stability region, and finite-temperature analyses~\cite{Holman:1992rv} suggest that thermal effects only enlarge it.}

\rev{The high formation scale makes the most direct external constraints
cosmological rather than laboratory based. In particular, the thermal
realisation is conditional on post-inflationary reheating populating the dark
sector above $T_{\rm nuc}\simeq(4$--$6)\times10^{14}$~GeV. This is below the
instantaneous-reheating upper limit inferred from the current tensor bound, but
it remains a non-trivial restriction on the inflationary scale, reheating
efficiency and dark-sector thermal history. Within the thermal realisation, additional probes can
arise directly from the decay of the dark-sector network. The finite string
segments and their monopole-like endpoints may release part of their energy
into the massless dark $\mathrm{U}(1)$ gauge boson or into massive dark
particles when they collapse and annihilate. Relativistic products would
contribute to dark radiation and hence to $\Delta N_{\rm eff}$ at BBN and
recombination~\cite{Nakai:2020oit,Planck:2018vyg}; stable massive products
could instead contribute to non-thermal dark matter, with a warm component
constrained by its free-streaming imprint on the Lyman--$\alpha$
forest~\cite{Irsic:2017ixq}. Quantitative predictions require the emission
spectra, branching fractions and subsequent dark-sector thermal history,
which we leave for future work.}

\rev{The full gravitational wave spectrum of quasi-stable string networks has
already been confronted with NANOGrav data in
Ref.~\cite{Lazarides:2023ksx}. The open task specific to this model is to
propagate the microscopic map
$(\ap,\sw,\beta)\mapsto(\Gmu,\kappa,T_{\rm nuc})$ through such a spectral fit,
rather than using the single-cutoff condition adopted here. Numerical modelling
of network scaling, friction and endpoint re-entry in the one-scale setup would
also be needed to predict the amplitude and detailed spectral shape beyond our
simplified network treatment.}

\section*{Acknowledgements}
We would like to thank Francesco Sannino for useful discussions.
\rev{In the preparation of this manuscript we made use of AI-assisted tools. Anthropic's Claude was employed for language editing, for cross-checking derivations and for assistance with figure preparation. OpenAI's Codex was employed for an independent re-derivation. All physical arguments, calculations and conclusions were developed and verified by the authors, who take full responsibility for them.} 

%%%==========================================================================
\appendix
%%%==========================================================================
\section{Average string length}
\label{app:string-length}

In this appendix, we derive the length distribution of strings and the associated average length as a function of their decay rate $\Gamma(T)$.

The starting point is a Boltzmann equation for the number density of segments $n(\ell, T)$ with length $\ell$ in an expanding Universe. The continuity equation for $n$ reads
\begin{equation}
    \partial_t n(\ell, T) + 3H\, n(\ell, T) = \text{(collision terms)}\,,
\end{equation}
where the $3H$ term accounts for Hubble dilution and ``$\text{(collision terms)}$'' accounts for all possible interactions between strings which change their number abundance. To rewrite this in terms of temperature, we use entropy conservation in radiation domination, which gives $T \propto a^{-1}$ and therefore $dT/dt = -HT$, $\partial_t = -HT\,\partial_T$. Substituting and dividing by $-HT$, the equation becomes
\begin{equation}
    \partial_T n(\ell, T) = \frac{3}{T}\, n(\ell, T) 
    - \frac{1}{HT}\,\text{(collision terms)}\,.
\end{equation}
Let us now focus on the collision term. We neglect loop formation or
gaining terms stemming from re-percolation, and consider only the
breaking term with rate per unit length per unit time, $\Gamma(T)$. The
collision term is then given by $-\Gamma(T)\,\ell\,n(\ell,T)$, where
$\Gamma(T)\,\ell$ is the total breaking rate of a segment of length
$\ell$ and accounts for the higher breaking probability for longer
segments.
The equation then reduces to
\begin{equation}
\label{eq:boltzmann}
    \partial_T n(\ell, T) = \frac{3}{T}\, n(\ell, T) 
    + \frac{\Gamma(T)\,\ell}{H(T)\,T}\, n(\ell, T)\,.
\end{equation}
The solution can be computed analytically and, up to a constant prefactor which depends on the initial conditions, it reads
\begin{equation}
    n(\ell, T) \propto T^3 \exp\left(-\ell\int_{T}^{T_{\rm nuc}} 
    \frac{\Gamma(T')}{H(T')\,T'}\, dT'\right)\propto \exp\left[-\ell/\bar{\ell}(T)\right]\,,
\end{equation}
which is an exponential distribution in $\ell$ with characteristic temperature-dependent length
\begin{equation}
\label{eq:ell_general}
    \bar{\ell}(T) \equiv \left(\int_{T}^{T_{\rm nuc}} 
    \frac{\Gamma(T')}{H(T')\,T'}\, dT'\right)^{-1}\,.
\end{equation}

The solution of Eq.~\eqref{eq:ell_general} strongly depends on the temperature dependence of the decay rate $\Gamma(T)$.
It is instructive to consider two limiting cases, which are discussed in the main text:
\begin{enumerate}
    \item First, if $\Gamma(T)=\Gamma_0$ is constant, in radiation
domination $H \sim T^2/M_{\rm Pl}$ and given $T_\nuc\gg T$ the
integral evaluates to $\sim\Gamma/H(T)$, giving
\begin{equation}
\label{eq:ell_constant}
    \bar{\ell}(T\ll T_{\rm nuc}) \sim \frac{H(T)}{\Gamma_0}\,,
\end{equation}
which decreases as the Universe cools.
\item Second, if $\Gamma(T)$ is sharply peaked around $T_\nuc$, it can be 
approximated as $\Gamma(T) \simeq \Gamma_\nuc$ over a narrow range 
near $T_\nuc$ and negligible elsewhere. The integral then localizes its argument around $T_\nuc$, giving
\begin{equation}
\label{eq:ell_star_app}
    \bar{\ell}(T\ll T_{\rm nuc}) \sim \frac{H_\nuc}{\Gamma_\nuc}\,.
\end{equation}
Therefore, below $T_\nuc$, the length distribution freezes, so that $\bar{\ell}(T)$ remains fixed at $\bar{\ell}(T_\nuc)$ as the Universe cools further. Since the thermal breaking rate of interest below is sharply localised near $T_\nuc$, Hubble stretching during the breaking epoch is negligible and may be omitted here. 
The subsequent stretching of the endpoint separation between $T_\nuc$ and $T_d$ is included explicitly at a later stage through the scale-factor factor in Eq.~\eqref{eq:Td_mech} in the main text.
\end{enumerate}

%%%==========================================================================
\section{Finite-temperature effective potential}
\label{app:thermal_potential}

This appendix collects the explicit field-dependent masses, the
renormalisation conditions, the thermal integral and the Debye
self-energies entering the effective potential of
Eq.~\eqref{eq:Veff_full}. We work along the neutral Higgs direction defined in
Eq.~\eqref{eq:phi_param}, along which $\Phi^\dagger\Phi = \phi^2/2$, so
the doublet potential of Eq.~\eqref{eq:model_lagrangian} reduces to
\begin{equation}
  V_{\rm tree}(\phi)
  \;=\; \frac{\lambda}{4}\left(\phi^2-\eta^2\right)^{\!2}.
  \label{eq:Vtree_app}
\end{equation}
The radial scalar mode $h$ and the three Goldstone modes $\chi$ have
field-dependent masses
\begin{equation}
  M_\Phi^2(\phi) \;=\; \lambda\!\left(3\phi^2-\eta^2\right)\,,
  \qquad
  m_\chi^2(\phi) \;=\; \lambda\!\left(\phi^2-\eta^2\right)\,,
  \label{eq:scalar_masses_app}
\end{equation}
which satisfy $M_\Phi^2(\eta)=2\lambda\eta^2$ and $m_\chi^2(\eta)=0$ as
expected. The field-dependent gauge-boson masses are
\begin{equation}
  M_W^2(\phi) \;=\; \frac{g^2\phi^2}{4}\,,
  \qquad
  M_Z^2(\phi) \;=\; \frac{(g^2+g'^2)\phi^2}{4}\,,
  \qquad
  m_A^2(\phi) \;=\; 0\,,
  \label{eq:gauge_masses_app}
\end{equation}
with multiplicities
\begin{equation}
  n_h = 1\,,
  \qquad
  n_\chi = 3\,,
  \qquad
  n_W = 6\,,
  \qquad
  n_Z = 3\,,
  \qquad
  n_\gamma = 2\,,
  \label{eq:dofs_app}
\end{equation}
where the gauge boson counts include all three polarisations. The photon-like neutral mode contributes only the two transverse
polarisations to $V_T$ (its longitudinal Debye state is dealt with separately below).
The one-loop zero-temperature contribution in Landau gauge is~\cite{Coleman:1973jx}
\begin{equation}
  V_{\rm CW}(\phi)
  \;=\;
  \sum_i\frac{n_i}{64\pi^2}\,
  m_i^4(\phi)\!\left[\log\frac{m_i^2(\phi)}{\mu_R^2}-c_i\right]\,,
  \qquad
  c_{\rm scalar} = \tfrac{3}{2}\,,
  \qquad
  c_{\rm gauge}  = \tfrac{5}{6}\,,
  \label{eq:VCW_app}
\end{equation}
in the $\overline{\rm MS}$ scheme. We fix the renormalisation scale to
$\mu_R = M_Z(\eta) = \tfrac12\sqrt{g^2+g'^2}\,\eta$, the natural scale of
the broken phase. Tachyonic contributions from scalar modes with
$m_i^2(\phi)<0$ at small $\phi$ -- the Goldstones for $\phi<\eta$ and
the radial Higgs for $\phi<\eta/\sqrt{3}$ -- are handled by replacing
$m^2$ with $|m^2|$ in the logarithm. This is the standard prescription,
and we have verified that the resulting imaginary part is numerically
negligible near the field configurations relevant for nucleation.

Adding the one-loop CW term in general shifts the position of the tree-level
minimum and the Higgs pole mass. We absorb both shifts by adding finite
counterterms
\begin{equation}
  V_{\rm ct}(\phi)
  \;=\; \frac{\delta\mu^2}{2}\phi^2
       +\frac{\delta\lambda}{4}\phi^4
       +\delta V_0\,,
  \label{eq:Vct_app}
\end{equation}
chosen to preserve the tree-level vacuum, scalar mass, and potential
normalisation,
\begin{equation}
  \left.\frac{\partial}{\partial\phi}
        \!\left(V_{\rm CW}+V_{\rm ct}\right)\right|_{\phi=\eta}\!=0\,,
  \qquad
  \left.\frac{\partial^2}{\partial\phi^2}
        \!\left(V_{\rm CW}+V_{\rm ct}\right)\right|_{\phi=\eta}\!=0\,,
  \qquad
  \left.\!\left(V_{\rm CW}+V_{\rm ct}\right)\right|_{\phi=\eta}\!=0\,.
  \label{eq:ct_conditions_app}
\end{equation}
With $d_1 \equiv V_{\rm CW}'(\eta)$ and $d_2\equiv V_{\rm CW}''(\eta)$,
the three conditions of Eq.~\eqref{eq:ct_conditions_app} fix the
counterterm coefficients to
\begin{equation}
  \delta\lambda \;=\; \frac{d_1/\eta - d_2}{2\eta^2},
  \qquad
  \delta\mu^2 \;=\; -\frac{d_1}{\eta} - \delta\lambda\,\eta^2,
  \qquad
  \delta V_0 \;=\; -V_{\rm CW}(\eta)
                   -\tfrac12\delta\mu^2\eta^2
                   -\tfrac14\delta\lambda\,\eta^4.
  \label{eq:ct_solution_app}
\end{equation}
This on-shell-inspired prescription is standard in studies of
finite-temperature symmetry breaking~\cite{Anderson:1991zb,Quiros:1999jp}.
In practice $\delta\lambda$ and $\delta\mu^2$ are small over most of the
strip but reach $\mathcal{O}(10\%)$ relative shifts at large $\alpha'$
and small $\beta$, where the radiative shift of the Higgs mass would
otherwise be non-negligible.

The bosonic finite-temperature contribution is obtained from the imaginary-time formalism~\cite{Matsubara:1955ws,Dolan:1973qd},
\begin{equation}
  V_T(\phi,T)
  \;=\;
  \frac{T^4}{2\pi^2}\sum_i n_i\,
  J_B\!\left(\frac{m_i^2(\phi)}{T^2}\right)\,,
  \qquad
  J_B(y) \;=\; \int_0^\infty\!\! dx\,x^2
  \log\!\left(1-e^{-\sqrt{x^2+y}}\right)\,,
  \label{eq:VT_app}
\end{equation}
where the sum runs over all bosonic modes of
Eq.~\eqref{eq:dofs_app}. The integral $J_B(y)$ is evaluated by direct
numerical integration of the exact thermal integral rather than via its
high-temperature expansion, to preserve accuracy down to $T\sim m_i$
where the asymptotic series loses precision. For negative $y$ (tachyonic regions of the
$V_{\rm eff}$ landscape) we analytically continue $J_B$ using the
prescription of Ref.~\cite{Quiros:1999jp}. The contribution from
tachyonic Goldstones is small near the relevant minima.
For the daisy resummation we use the scheme of
Ref.~\cite{Arnold:1992rz}; the corresponding contribution to the
effective potential reads
\begin{equation}
  V_{\rm daisy}(\phi,T)
  \;=\;
  -\frac{T}{12\pi}\sum_{j\in\mathcal{R}}\,n_j\!\left[
    \bigl(\,m_j^2(\phi)+\Pi_j(T)\,\bigr)^{3/2}
    -\bigl(m_j^2(\phi)\bigr)^{3/2}\right]\,,
  \label{eq:Vdaisy_app}
\end{equation}
where the resummed set
$\mathcal{R}=\{W_L^\pm,Z_L,\gamma_L,h,\chi\}$ comprises the longitudinal
gauge zero modes (with $n_{W_L^\pm}=2$ and $n_{Z_L}=n_{\gamma_L}=1$) and
the scalar modes (with $n_h=1$ and $n_\chi=3$). The transverse gauge
polarisations are not infrared sensitive and do not require resummation.
For the charged $W_L^\pm$ and the scalar modes the resummed mass is
$m_j^2(\phi)+\Pi_j(T)$ as written. For the neutral longitudinal modes
$Z_L,\gamma_L$, the resummed masses are the eigenvalues of the
Debye-corrected $(W^3,B)$ matrix introduced in
Eq.~\eqref{eq:neutral_debye_matrix_app} below, and the unresummed
subtraction $\bigl(m_j^2(\phi)\bigr)^{3/2}$ is constructed from the
corresponding $T=0$ eigenvalues $M_Z^2(\phi)$ and $0$.
The Debye self-energies in this dark $\mathrm{SU}(2)\times\mathrm{U}(1)$
sector, with one complex Higgs doublet and no fermions, are
\begin{equation}
  \Pi_W(T) \;=\; \Pi_{W^3}(T) \;=\; \frac{5\,g^2 T^2}{6},
  \qquad
  \Pi_B(T) \;=\; \frac{g'^2 T^2}{6}\,,
  \label{eq:Pi_gauge_app}
\end{equation}
\begin{equation}
  \Pi_h(T) \;=\; \Pi_\chi(T)
            \;=\; T^2\!\left(\frac{3 g^2}{16}+\frac{g'^2}{16}
                              +\frac{\lambda}{2}\right)\,.
  \label{eq:Pi_scalar_app}
\end{equation}
The $5/6$ factor in $\Pi_W$ reflects the SU(2) gauge self-loop
($C_A/3 = 2/3$) together with the Higgs-doublet contribution, which in
our generator normalisation $T^a=\tau^a/2$ for a single complex
fundamental doublet evaluates to $1/6$. The abelian $\Pi_B$ receives
only the Higgs-doublet contribution since there is no
$B$-self-interaction~\cite{Arnold:1992rz}.

The longitudinal eigenmasses entering Eq.~\eqref{eq:Vdaisy_app} are
obtained by adding the Debye corrections in the $(W^3,B)$ basis before
diagonalising. With the field-dependent gauge masses written as a
$2\times 2$ matrix
\begin{equation}
  \mathcal{M}^2(\phi,T)
  \;=\;
  \begin{pmatrix}
    \tfrac14 g^2\phi^2 + \Pi_{W^3}(T) & -\tfrac14 g g'\phi^2 \\[2pt]
    -\tfrac14 g g'\phi^2 & \tfrac14 g'^2\phi^2 + \Pi_B(T)
  \end{pmatrix},
  \label{eq:neutral_debye_matrix_app}
\end{equation}
the longitudinal eigenmasses $m_{Z_L}^2(\phi,T)$ and
$m_{\gamma_L}^2(\phi,T)$ are the two eigenvalues of $\mathcal{M}^2$, and
the longitudinal charged $W$ mass is simply
$m_{W_L}^2(\phi,T) = \tfrac14 g^2\phi^2 + \Pi_W(T)$. In tachyonic
regions, where the resummed masses become negative, we retain the real
part of the daisy contribution, consistently with the prescription used
for the thermal function $J_B$.

For each point $(\alpha',\sin^2\theta_w,\sqrt{\beta})$ of the
parameter-space scan, we assemble $V_{\rm eff}(\phi,T)$ via
Eq.~\eqref{eq:Veff_full} with the inputs above. The potential is
shifted at every temperature by a $\phi$-independent constant so that
$V_{\rm eff}(0,T)=0$. We then locate the broken minimum $\phi_+(T)$ by
gradient descent on the radial slice; determine the critical temperature
$T_\star$ by demanding degeneracy
$V_{\rm eff}(\phi_+(T_\star),T_\star)=V_{\rm eff}(0,T_\star)$; compute
$S_3(T)$ on a two-stage temperature grid (dense near $T_\star$, coarser
for $T<0.9\,T_\star$) using the O(3)-symmetric solver of
\texttt{CosmoTransitions}~\cite{Wainwright:2011kj}.

%%%==========================================================================
\bibliographystyle{JHEP}
\bibliography{refs}

\end{document}